# On-Chip Test Infrastructure Design for Optimal Multi-Site Testing of System Chips

Sandeep Kumar Goel     Erik Jan Marinissen

Philips Research Laboratories
IC Design – Digital Design & Test
Prof. Holstlaan 4, M/S WAY-41
5656 AA Eindhoven, The Netherlands

{SandeepKumar.Goel, Erik.Jan.Marinissen}@philips.com

**Abstract**

*Multi-site testing is a popular and effective way to increase test throughput and reduce test costs. We present a test throughput model, in which we focus on wafer testing, and consider parameters like test time, index time, abort-on-fail, and contact yield. Conventional multi-site testing requires sufficient ATE resources, such as ATE channels, to allow to test multiple SOCs in parallel. In this paper, we design and optimize on-chip DfT, in order to maximize the test throughput for a given SOC and ATE. The on-chip DfT consists of an E-RPCT wrapper, and, for modular SOCs, module wrappers and TAMs. We present experimental results for a Philips SOC and several* ITC'02 SOC Test Benchmarks.

## 1   Introduction

The manufacturing test costs of subsequent generations of SOCs threaten to increase beyond what is acceptable, if no proper countermeasures are taken. Factors that drive the (digital) test costs up are the increases in pin count, test data volume, speed and corresponding required ATE accuracy. Especially the test data volume has risen dramatically, due to a combination of growth in transistor count and new advanced test methods (such as delay-fault testing) which add significantly to the test set size. As a consequence, testing of 'monster chips' [1] requires expensive ATEs with a large channel count and deep test vector memory [2].

Several methods are applied to reduce the test costs. With Built-In Self Test (BIST), SOCs test (parts of) themselves and hence eliminate the need for ATE altogether. However, BIST is expensive to implement on-chip, and hence its usage is typically limited to applications which require in-field testing. Test Data Compression (TDC) techniques still require the presence of an ATE, but reduce the demands on both vector memory and test application time by exploiting the many *don't care* bits in the test set to compress it. Another effective approach to reduce test cost, orthogonal to TDC, is *multi-site* testing, in which multiple instances of the same SOC are tested in parallel on a single ATE [3, 4, 5, 6, 7]. More sites mean more devices are tested in parallel. Multi-site testing amortizes the fixed ATE costs over multiple SOCs.

Efficient multi-site testing requires effective management of test resources as the number and depth of ATE channels and the on-chip DfT, while taking into account parameters such as prober index time, contact yield, etc. One way to allow an increase in the number of sites is to increase the number of ATE channels. However, this solution not only brings substantial extra costs, but also is not very scalable for SOCs with high pin counts. The other way to increase the number of sites is to narrow down the SOC-ATE interface, i.e., the number of SOC terminals that needs to be contacted during testing. Reduced-Pin-Count-Test (RPCT) [8, 9, 4] is a well-known DfT technique that does exactly this.

This paper focuses on designing and optimizing an on-chip test infrastructure (DfT) to facilitate high-throughput multi-site wafer testing of large SOCs. We assume a given and fixed target *test cell*, including ATE and probe station. We present a two-step algorithm that designs the SOC test infrastructure such that the SOC test data volume fits on the target ATE within a single load and the multi-site test throughput is maximum. The number of sites at which the throughput is maximum, we refer to as 'optimal multi-site'. This is a different optimization criterion from simply maximizing the number of multi-sites; a large number of sites means less ATE channels per SOC, which in turn increases the test application time per SOC. Consequently, in order to minimize test costs through multi-site testing, the number of sites should be tuned such that the test throughput is maximized. In case the SOC uses a flattened top-level test, the algorithm determines all parameters to design an Enhanced-RPCT wrapper [9]. In case the given SOC uses a modular, core-based test approach [10], in addition to the E-RPCT wrapper, the algorithm determines the on-chip test architecture consisting of TAMs and core wrappers [11, 12].

The outline of this paper is as follows. Section 2 reviews the prior work in this domain. Section 3 describes our assumptions regarding multi-site testing. Section 4 details our cost model for test time and test throughput. Section 5 formally defines the problem of designing on-chip test infrastructures for optimal multi-site testing of SOCs, while Section 6 describes our two-step algorithm to solve it. Section 7 contains experimental results for the Philips SOC PNX8550 [1] and several SOCs taken from the *ITC'02 SOC Test Benchmarks* [13]. Finally, Section 8 concludes the paper.

## 2   Prior Work

Reduced-Pin-Count Testing (RPCT) is a DfT technique used to reduce the number of IC pins that need to be contacted by the ATE. RPCT assumes the presence of internal- and boundary-scan. The basic principle of RPCT is that only the input and output terminals of the scan chains (including the boundary-scan chain), test control pins, and clock pins need to be connected to ATE channels. Access to all other functional pins is achieved via the boundary-scan chain. First use of RPCT with LSSD boundary-scan was re-ported by IBM to enable the use of low-cost ATE for ASICs [8]. Since then, several extensions have been made to the basic RPCT technique. Two such extensions are re-configurable RPCT [4] and Enhanced RPCT [9]. In [4], a technique to design an RPCT wrapper around an SOC is presented. In this case, the DfT for the SOC can be designed without even knowing about the target ATE. Later, by using the re-configurable logic, the number of scan chains and their length can be modified in the RPCT wrapper according to the ATE specification. The basic idea behind E-RPCT [9] is to provide access not only to the functional terminals, but also to the internal scan chains via the boundary-scan architecture, in order to enable even further scalability of the SOC-ATE interface. The E-RPCT wrapper around an IC truly converts $n$ external test inputs and outputs into $m$ internal test inputs and outputs, for all integers $n, m$ with $0 < n < m$. Our paper is based on the usage of E-RPCT.

Many papers published in the domain of multi-site testing model the economics of multi-site testing for test cost reduction. In [3], a conceptual semiconductor test-economic model is used to analyze the interrelations between the parameters that make up the test cost per device, such as test time, index time, yield, utilization, ATE capital cost, etc. In [5, 6], similar but rather limited cost models are described. The only paper that presents techniques to design and optimize on-chip test hardware to enable multi-site testing is [7]. It presents a rectangle bin-packing based technique to design the test architecture (consisting of TAMs and core wrappers) for a core-based SOC with a target ATE, such that the test architecture requires a minimum number of ATE channels and the SOC test data volume fits on the given ATE. A minimum number of ATE channels per device enables the maximum multi-site testing possible for the given SOC. While this paper was the first one in this domain, it has several limitations. The paper only discusses the design of core wrappers and TAMs for modularly tested, core-based SOCs, and ignores the design of chip-level RPCT wrappers. It maximizes the number of sites that can be tested in parallel, while we show in our paper that this does not always yield maximum throughput. To maximize the number of sites, the paper assumes that a common set of input channels can be used to broadcast test stimuli to all sites, which is often not practical. And finally, [7] looks at the test time only, and does not take into account parameters like index time, contact yield and re-test rate, and abort-on-fail.

## 3  Multi-Site Test Flow

Multi-site testing can be done at wafer test as well as at final ('packaged IC') test. In this paper, we assume the following two-step test flow.

1. During *wafer test*, the internal circuitry of the SOC die in question is tested. This is done through a narrow E-RPCT interface, in order to enable a large number of multi-sites, as well as to reduce the chances for contact test fails. The non-E-RPCT pins of the SOC are not contacted.

2. During *final test*, the IOs of the packaged SOC are tested. For this purpose, *all* pins of the SOC are contacted. Optionally, the SOC internal circuitry can be tested again, either through all pins, or through their E-RPCT subset. The number of multi-sites during final test is limited by the number of available ATE channels divided by the number of pins per SOC, as well as by the available device handler.

In this paper, we focus on maximizing the multi-site test throughput during wafer testing. The complete wafer is located near the ATE, and the E-RPCT bonding pads of the SOCs under test are physically probed. Today's high channel-count ATEs and corresponding high pin-count probe technologies enable massive multi-site testing. Unfortunately, the circular shape of the wafer brings some losses in multi-site testing at the periphery of the wafer; these are ignored in the sequel of this paper.

Iyengar et al. [7] assumed that ATE stimuli are broadcasted to all multi-sites. However, stimuli broadcast is often impractical. Some ATEs simply do not support broadcasting; they assign a channel to a site, and, if that site fails, no more stimuli will be sent to the device under test. Furthermore, broadcasting can cause undesired side effects, such as a fault at the bonding pad of one site causing incorrect test results on other sites. Hence, our solution approach in this paper explicitly supports both the cases *with* and *without* stimuli broadcasting. Obviously, the optimal multi-site for the case without stimuli broadcast is significantly lower than for the case with stimuli broadcast.

## 4  Multi-Site Cost Model

The total time spent on a set of devices to be tested in parallel is the sum of the index time $t_i$ and the test application time $t_a$, as depicted in Figure 1.

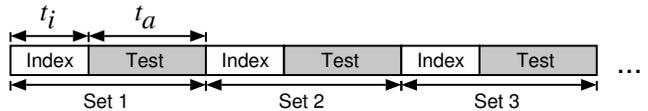

**Figure 1:** Wafer testing time consists of index time $t_i$ and test time $t_a$.

The *index time* $t_i$ is the time required to position the probe interface in order to make contact with the bonding pads of the SOC(s) under test. We assume the index time to be a constant, dependent on the type of probe station. A typical value is $t_i = 0.7$ s.

The test consists of a contact test and a manufacturing test. In the *contact test*, it is checked whether all terminals required for the subsequent manufacturing test are properly connected to the ATE. If one or more of these terminals are not properly connected, the SOC fails the contact test. The probability $p_c$ of a single terminal to pass the contact test, also referred to as the *contact yield*, needs to be high to be able to successfully test high pin-count SOCs. All terminals undergo their contact test simultaneously, and hence the contact test time $t_c$ is a constant. A typical value is $t_c = 10$ ms.

During the *manufacturing test*, the SOC is checked for manufacturing faults. In this paper, we only consider digital tests, i.e., for logic and memories. The probability $p_m$ of a single SOC to pass the manufacturing test is also referred to as the *yield*. The manufacturing test time $t_m$ depends on the width of the RPCT test interface, the test data volume, and how well the various SOC tests can be scheduled.

The total test time can now be written as

$$t = t_i + t_a = t_i + t_c + t_m. \qquad (4.1)$$

In high-volume production testing, where faulty devices are often not analyzed, but simply discarded, it is possible to abort the test as soon as the first failing test vector is observed. This *abort-on-fail* strategy can significantly reduce the average test time per device, especially in the case of relatively low yields. As we will show in Section 7, multi-site testing reduces the effect of the abort-on-fail strategy on the average test time, as now tests can only be aborted if all $n$ sites have started failing, which is simply less likely to happen. For an SOC with $k$ pins involved in its test, the probability $P_c$ that at least one out of $n$ SOCs will pass the contact test is

$$P_c = 1 - (1 - p_c{}^k)^n. \quad (4.2)$$

Similarly, the probability $P_m$ that at least one out of $n$ SOCs will pass the manufacturing test is

$$P_m = 1 - (1 - p_m)^n. \quad (4.3)$$

Based on the assumption that failing SOCs do not take any test time ($t_m = 0$), a theoretical lower bound on the total test application time for a set of $n$ devices can be written as

$$t_a = t_c + P_c \times P_m \times t_m. \quad (4.4)$$

The choice for the obviously unrealistic assumption that $t_m = 0$ for failing SOC is motivated by the fact that it allows us to make a strong conclusion about the reduced effectiveness of abort-on-fail in multi-site testing in Section 7.

Assuming a full utilization of the ATE, the total number of devices tested per hour $D_{\text{th}}$ for $n$ multi-site testing can be written as

$$D_{\text{th}} = \frac{3600 \times n}{t_i + t_a}. \quad (4.5)$$

In this equation, both $t_i$ and $t_a$ are in seconds. Furthermore, $t_a$ can be either the original test application time, or, when abort-on-fail is used, the reduced test application time of Equation 4.4.

In many companies, it is common practice to re-test those devices that failed only on their contact test. The premise of re-testing is that the chances are high that the failure was caused by a wrong probe contact, rather than that the SOC itself was faulty. If that is indeed the case, it would be a waste to discard basically good products. Excluding the unlikely event of multiple failing terminal contacts per SOC, $D_{\text{th}}$ SOCs with $k$ terminals each and a contact yield $p_c$ per terminal, will require $(1 - p_c) \times k \times D_{\text{th}}$ SOCs per hour to be re-tested. While the number of devices tested per hour $D_{\text{th}}$ remains unaffected, re-testing has an impact on the number of unique devices tested per hour $D_{\text{th}}^u$. Assuming at most one failing terminal contact per SOC, and assuming that devices are re-tested at most once, $D_{\text{th}}^u$ can be written as

$$D_{\text{th}}^u = (1 - (1 - p_c) \times k) \times D_{\text{th}}. \quad (4.6)$$

## 5 Problem Definitions

The problem of test infrastructure design for optimal multi-site testing (i.e., with maximal throughput) of modularly-tested (core-based) SOCs can be formally defined as follows.

**Problem 1** [Optimal Multi-Site Testing for Core-Based SOCs]
Given an SOC consisting of a set of modules $M$, and for each Module $m \in M$ the number of test patterns $p(m)$, the number of functional input terminals $i(m)$, the number of functional output terminals $o(m)$, the number of functional bidirectional terminals $b(m)$, the number of scan chains $s(m)$, and for each scan chain $j$, the length of the scan chain in flip flops $l(m, j)$. Given a target ATE with $N$ channels, each with vector memory depth $V$. Furthermore is given a target SOC probe station with index time $t_i$. Determine the number of multi-sites $n$, the number of ATE channels per site $k$ ($k$ even), and a $k/2$-to-$w$ E-RPCT wrapper for the SOC and test architecture (i.e., determine the number of TAMs, the width of these TAMs, the assignment of modules to TAMs, and the core wrapper design per module [11, 12]), resulting in $T$ test clock cycles per SOC, such that during $n$ multi-site testing

1. $n \times k \leq N$, i.e., the number of required ATE channels does not exceed the number of available ATE channels;
2. $T \leq V$, i.e., the required ATE vector memory depth does not exceed the available depth;
3. the test throughput $D_{\text{th}}$ is maximum. □

DfT solutions to Problem 1 look like the architecture depicted in Figure 2(a). Problem 1 addresses the case of a modularly-tested (core-based) SOC. Problem 2, in which the same problem is solved for a flattened SOC, actually turns out to be a degenerate case of Problem 1. For a flattened SOC, we simply deal with one module, i.e., $|M| = 1$. The module wrapper and E-RPCT wrapper coincide, and there are no TAMs. This case is depicted in Figure 2(b).

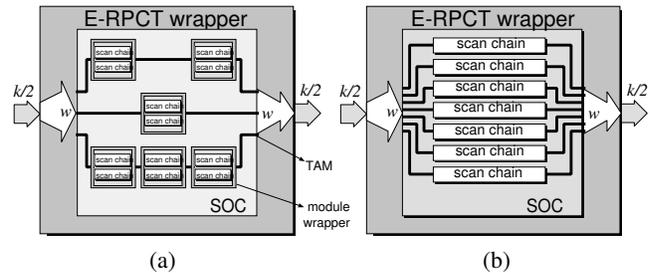

**Figure 2:** Test infrastructure design for (a) modular and (b) flat SOCs.

Problems 1 and 2 actually come in several variants: (1) without stimuli broadcast ($n \times k \leq N$) and with stimuli broadcast ($n \times \frac{k}{2} + \frac{k}{2} \leq N$), (2) without abort-on-fail ($t_a = t_c + t_m$) and with abort-on-fail ($t_a = t_c + P_c \times P_m \times t_m$), and (3) without re-test (maximizing $D_{\text{th}}$) and with re-testing (maximizing $D_{\text{th}}^u$).

## 6 On-Chip Infrastructure Design

In this section, we present a two-step algorithm that solves Problems 1 and 2. In Step 1 of the algorithm, we determine the maximum multi-site $n_{\max}$ for the given SOC and ATE, and the corresponding test infrastructure. In Step 2, we use linear search to find the number of sites $n_{\text{opt}}$ ($1 \leq n_{\text{opt}} \leq n_{\max}$) for which the test throughput is maximum and we modify the test infrastructure accordingly. Details of both steps are given below.

**Step 1**: While determining $n_{\max}$, we use two optimization criteria, as illustrated in Figure 3(a). Criterion 1 is the minimization of the number of ATE channels $k$ utilized by one SOC, such that the test still fits into the vector memory depth $V$ of the ATE. Criterion 2 is the minimization of the actual filling of the vector memory. Criterion 1 has priority, as it maximizes the number of sites, as shown in Figure 3(b). Criterion 2 is meant to reduce the test application time per SOC.

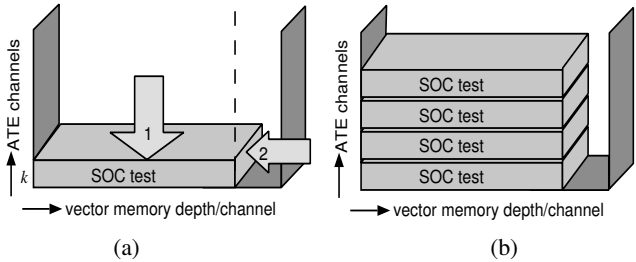

**Figure 3:** Fitting SOC test data on the target ATE with as few ATE channels as possible in order to allow the maximum number of multi-sites.

In this step, we first calculate the minimum number of ATE channels $k_{\min}(m)$ required for every module $m \in M$ such that module's test time does not exceed the ATE vector memory depth per channel $V$. To design the wrapper around a module for a given number of ATE channels, we use the COMBINE algorithm presented in [14]. If $k_{\min}(m) > N$, for any $m \in M$, then the SOC cannot be tested on the target ATE and the procedure is exited. Otherwise, modules are sorted in decreasing order of their $k_{\min}(m)$. Now we start with the first module, assign it the $k_{\min}(m)$ number of channels and form a *channel group* of width $k_{\min}(m)$. Iteratively, we move to the next module and check whether it can be assigned to an already existing channel group without exceeding its vector memory depth limit. If more than one such channel groups are found, then the module is assigned to the group that requires the smallest vector memory depth.

In case, the module cannot be assigned to any existing channel group, we consider two options: (1) create a new channel group, or (2) increase the width of an existing channel group such that the module can be assigned without exceeding the vector memory limit. We select the best of the two options, i.e., the option in which the total free memory available on all used channels is maximum. This minimizes the test application time for the SOC considering the same number of channels. In both options, we take into account that the total number of used channels does not exceed $N$. If the assignment of a module leads to the violation of this constraint, then the SOC cannot be tested on the ATE and the procedure is exited. The procedure is repeated until all modules are assigned.

The summed width of all channel groups determines the total number of channels $k$ for the SOC. Similarly, the test application time for the SOC is equal to the maximum of the filled vector memory over all channel groups. The maximum multi-site possible $n_{\max} = \lfloor \frac{N}{k} \rfloor$ in the case without stimuli broadcast, and $n_{\max} = \lfloor \frac{2N}{k} \rfloor - 1$ in the case with stimuli broadcast.

One iteration of Step 1 is illustrated in Figure 4. Figure 4(a) shows a situation in which Cores $A$ and $B$ have already been assigned to a TAM that requires $k1$ ATE channels, while Core $C$ is assigned to another TAM that requires $k2$ ATE channels. In Figure 4(b), the algorithm tries to add Core $D$ to either one of the two already existing TAMs. Unfortunately, both alternatives exceed the vector memory depth limit $V$. Hence, the algorithm is forced to start using more ATE channels in order to add Core $D$. In Figure 4(c), the three alternatives considered are depicted. Alternative (i) is to add a new channel group for Core $D$, in this case with $k3 = k_{\min}(D)$ ATE channels. Alternative (ii) extends TAM 1 from $k1$ to $k1 + k3$ channels, and is only valid if Core $D$ can now be added without exceeding the vector memory depth $V$. Similarly, Alternative (iii) extends TAM 2 from $k2$ to $k2 + k3$ channels. The alternative which yields the smallest vector memory filling is selected.

**Step 2**: In this step, we identify the number of sites $n_{\text{opt}}$ for which the throughput $D_{\text{th}}$ is maximum. We use linear search from $n_{\max}$ down to 1 to calculate the corresponding $D_{\text{th}}$ value. In every iteration, we try to redistribute the ATE channels $k_{\text{free}}$ freed up by giving up one site over the remaining sites. Only if $k_{\text{free}} > 2n$ (for the case without stimuli broadcast) or $k_{\text{free}} > n + 1$ (for the case with stimuli broadcast), redistribution makes sense. In so, for each site, we assign iteratively free channels to the channel group that is maximally filled. This can reduce the test application time per site. We record the throughput for the value of $n$. Finally, after the linear search, we find $n_{\text{opt}}$ as the number of sites for which the throughput $D_{\text{th}}$ is maximum.

Figure 5 illustrates the operation of the proposed algorithm for the Philips SOC PNX8550 [1], for both the cases with and without stimuli broadcast. For the target ATE, we assumed $N = 512$ channels and $V = 14$ M vector memory per channel. Furthermore, we consider 5 MHz clock speed for the test clock, an index time $t_i = 0.7$ s, and a contact test time $t_c = 10$ ms. For the case *without* broadcast, Step 1 already yielded the optimal result, i.e., $n_{\max} = n_{\text{opt}} = 15$, and the corresponding throughput is $D_{\text{th}} = 15,780$ devices per hour. However, for the case *with* broadcast, Step 1 results in $n_{\max} = 24$ multi-sites, with a throughput $D_{\text{th}} = 27,270$ devices per hour, whereas Step 2 finds $n_{\text{opt}} = 21$, with a corresponding maximum throughput $D_{\text{th}} = 28,062$ devices per hour. At $n_{\max} = 24$, only 40 channels per site are used. As we decrease the number of sites, channels get freed-up. Initially there are insufficient freed-up channels to be able increase the channel width to the remaining sites, i.e., $k_{\text{free}} < n + 1$, and hence, throughput $D_{\text{th}}$ only decreases. $D_{\text{th}}$ starts to increase again at $n = n_{\text{opt}} = 21$. The straight, dashed line shows, again for the stimuli broadcast case, which throughputs would have been obtained for various multi-sites, based on Step 1 only. If for some reason (e.g., equipment), the multi-site is limited to, say, $n \leq 16$, Steps 1+2 together result in 34% more throughput than Step 1 only.

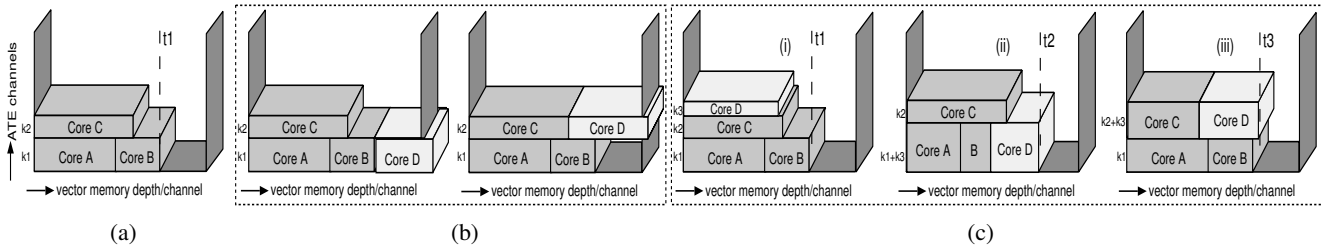

**Figure 4:** Example illustration of Step 1 for an SOC with two cores.

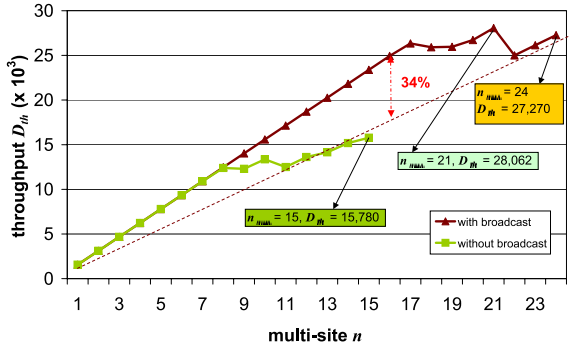

**Figure 5:** Example illustrating the operation of the proposed algorithm for Philips SOC PNX8550.

## 7 Experimental Results

First, we present experimental results for the Philips SOC PNX8550 [1]. This modular, core-based SOC is based on the Philips Nexperia$^{\text{TM}}$ Home Platform and contains 62 logic and 212 memory modules. In our experiments, we assume $N = 512$, $V = 7$ M, a test clock of 5 MHz, no stimuli broadcast, $t_i = 0.7$ s, and $t_c = 10$ ms (unless specified otherwise).

Figure 6 shows what happens to the test throughput $D_{\text{th}}$ if we extend our basic ATE with more channels (Figure 6(a)) or deeper vector memory (Figure 6(b)). The figures illustrate that the test throughput increases linearly with the number of ATE channels; by doubling the number of ATE channels, the test throughput can be doubled. This is due to the fact that the number of sites increases linearly with the number of channels, while the test time remains constant. On the other hand, the test throughput does not increase linearly with the vector memory depth. This is due to the fact that an increase in test vector memory depth leads both to an increase in multi-site, as well as to an increase in test application time. Therefore, doubling the test vector memory does *not* result in a double throughput.

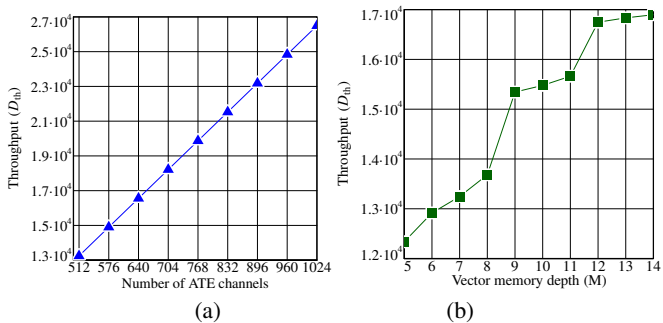

**Figure 6:** Variation in throughput with (a) number of ATE channels and (b) vector memory depth.

However, the cost of increasing the vector memory depth is rather small compared to the cost of increasing the number of ATE channels. According to standard market prices, buying 16 additional ATE channels with 7 M memory depth would cost roughly USD 8,000. At the same time, upgrading test vector memory for 16 channels from 7 M to 14 M would cost only USD 1,500. Therefore, if we double the test vector memory for all 512 channels, it will cost around $(512/16) \times 1500 =$ USD 48,000. For this money, the increase in test throughput is 27%. For the same amount of money, we can buy roughly 96 channels. This will result in a increase of 18% in test throughput. Therefore, for the same cost, increasing the test vector memory depth is more beneficial than increasing the number of ATE channels.

Subsequently, we analyze the impact of the re-test rate on the test throughput. Figure 7(a) shows the variation in the unique test throughput $D_{\text{th}}^u$ with the contact yield $p_c$. From the figure, we can see that the negative impact of re-testing on $D_{\text{th}}^u$ decreases with increasing vector memory depth. This is due to the fact that with deep test vector memory, less ATE channels are used per device, and hence the re-test rate is small. However, for small test vector memories, there is a significant drop in the number of unique devices per hour for low contact yield. Therefore, it can be concluded that deep test vector memory is not only useful from a test throughput point of view, but also from a contact yield point of view.

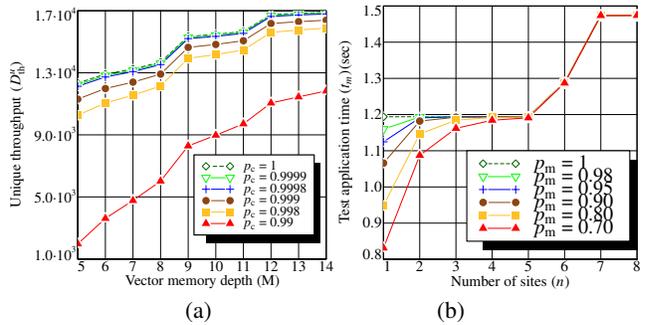

**Figure 7:** Impact of (a) re-testing of devices on the test throughput, and (b) abort-on-fail technique on total test time in a multi-site environment.

Next, we show the influence of multi-site testing on the effectiveness of applying the abort-on-fail technique. In Figure 7(b), we show the variation in the test time $t_m$ with SOC yield $p_m$. The figure shows that increased multi-site testing quickly reduces the positive effect of applying abort-on-fail. Even at a low yield of 70% (and under the overly optimistic assumption that for failing devices $t_m = 0$, see Equation 4.4), the effectiveness of abort-on-fail becomes invisible beyond $n = 5$.

Finally, we compare the results of the proposed algorithm to those published by Iyengar et al. [7] for several *ITC'02 SOC Test Benchmarks* [13]. [7] assumes stimuli broadcast and calculates $n_{\max}$ instead of $n_{\text{opt}}$. In order to compare on an equal basis, for this comparison we also assumed stimuli broadcast and have only applied Step 1 of our algorithm. Table 1 lists the number of ATE channels $k$ used for a single SOC. We report both a theoretical lower bound on $k$ from [7], as well as our result. In most cases, our algorithm matches the lower bound. The table also lists the maximum multi-site $n_{\max}$, obtained by [7] and by us. In all cases, except for SOC p93791 with 1.768 M channel depth, our algorithm obtains a higher multi-site.

| SOC d695 | | $N = 256$ | | | SOC p22810 | | $N = 512$ | | | SOC p34392 | | $N = 512$ | | | SOC p93791 | | $N = 512$ | | |
|---|---|---|---|---|---|---|---|---|---|---|---|---|---|---|---|---|---|---|---|---|
| $V$ | $k$ | | $n_{\max}$ | | $V$ | $k$ | | $n_{\max}$ | | $V$ | $k$ | | $n_{\max}$ | | $V$ | $k$ | | $n_{\max}$ | |
| | LB [7] | Us | [7] | Us | | LB [7] | Us | [7] | Us | | LB [7] | Us | [7] | Us | | LB [7] | Us | [7] | Us |
| 48K | 28 | 28 | 16 | 17 | 384K | 36 | 36 | 23 | 27 | 768K | 40 | 40 | 21 | 24 | 1.000M | 54 | 58 | 16 | 16 |
| 56K | 24 | 24 | 18 | 20 | 448K | 30 | 32 | 27 | 31 | 896K | 34 | 34 | 24 | 29 | 1.256M | 44 | 46 | 21 | 21 |
| 64K | 22 | 22 | 22 | 22 | 512K | 26 | 28 | 31 | 35 | 1.000M | 30 | 30 | 31 | 33 | 1.512M | 36 | 38 | 24 | 25 |
| 72K | 18 | 20 | 24 | 24 | 576K | 24 | 24 | 35 | 41 | 1.128M | 26 | 26 | 33 | 38 | 1.768M | 32 | 36 | 29 | 27 |
| 80K | 18 | 18 | 27 | 27 | 640K | 22 | 22 | 38 | 45 | 1.256M | 24 | 24 | 35 | 41 | 2.000M | 28 | 28 | 33 | 35 |
| 88K | 16 | 16 | 31 | 31 | 704K | 20 | 20 | 45 | 50 | 1.384M | 22 | 22 | 38 | 45 | 2.256M | 24 | 26 | 38 | 38 |
| 96K | 14 | 14 | 31 | 35 | 768K | 18 | 18 | 45 | 55 | 1.512M | 20 | 20 | 41 | 50 | 2.512M | 22 | 24 | 41 | 41 |
| 104K | 14 | 14 | 35 | 35 | 832K | 16 | 18 | 50 | 55 | 1.640M | 18 | 18 | 45 | 55 | 2.768M | 20 | 20 | 45 | 50 |
| 112K | 12 | 12 | 35 | 41 | 896K | 16 | 16 | 50 | 63 | 1.768M | 18 | 18 | 50 | 55 | 3.000M | 18 | 20 | 50 | 50 |
| 120K | 12 | 12 | 41 | 41 | 960K | 14 | 14 | 55 | 72 | 1.896M | 16 | 16 | 50 | 63 | 3.256M | 18 | 18 | 55 | 55 |
| 128K | 12 | 12 | 41 | 41 | 1M | 14 | 14 | 63 | 72 | 2.000M | 16 | 16 | 55 | 63 | 3.512M | 16 | 18 | 63 | 63 |

**Table 1:** Experimental results for maximum multi-site for the rectangle bin-packing algorithm in [7] and our new algorithm.

# 8 Conclusion

To reduce test cost, multi-site testing is an effective approach. In this paper, we considered multi-site wafer testing and modeled the test throughput considering parameters like test time, index time, stimuli broadcast, abort-on-fail, and contact yield. We showed that multi-site testing requires optimizing the design of the on-chip test infrastructure.

To design the test infrastructure for a given SOC with a fixed target ATE, we presented a two-step algorithm. We design the test infrastructure in such a way, that the SOC test data volume fits on the target ATE and the test throughput is maximum. For a given SOC, the presented technique determines the parameters required to design an E-RPCT wrapper around the SOC. If the given SOC is core-based, then the procedure also determines the on-chip test architecture consisting of core wrappers and TAMs.

We presented experimental results for Philips SOC PNX8550, as well as for several *ITC'02 SOC Test Benchmarks*. The results show that the proposed algorithm outperforms other published approaches. Experimental results also show that to increase the test throughput, increasing the vector memory depth is more cost-effective than increasing the number of ATE channels. Finally, we conclude that benefits of the abort-on-fail technique are rather limited when used in combination with multi-site testing.

**Acknowledgements**

We thank Peter Lagner of Advantest in Munich, Germany, Domenico Chindamo of Agilent Technologies in Rome, Italy, and Stefan Eichenberger and Robert van Rijsinge of Philips Semiconductors in Nijmegen, The Netherlands for fruitful discussions during the conception of this work.